\documentclass[runningheads]{llncs}
\usepackage{amsmath}
\usepackage{graphicx}
\usepackage{amssymb}
\usepackage{romannum}
\usepackage{color}
\usepackage{cite}
\usepackage[misc]{ifsym}
\begin{document}
\title{Detecting Pancreatic Ductal Adenocarcinoma in Multi-phase CT Scans via Alignment Ensemble}
%\title{Deep Alignments for Detecting Pancreatic Adenocarcinoma in Multi-phase CT Scans}
%
\titlerunning{Detecting PDAC in Multi-phase CT Scans via Alignment Ensemble}
% If the paper title is too long for the running head, you can set
% an abbreviated paper title here
%
%\author{First Author\inst{1}\orcidID{0000-1111-2222-3333} \and
%Second Author\inst{2,3}\orcidID{1111-2222-3333-4444} \and
%Third Author\inst{3}\orcidID{2222--3333-4444-5555}}
%\author{Anonymous Authors}
\author{Yingda Xia\inst{1}\textsuperscript{,*}, Qihang Yu\inst{1}\textsuperscript{,*}, Wei Shen\inst{1}$^{(\textrm{\Letter})}$, Yuyin Zhou\inst{1}, \\ Elliot K. Fishman\inst{2} \and Alan L. Yuille\inst{1}}
% index{Xia, Yingda}
% index{Yu, Qihang}
% index{Shen, Wei}
% index{Zhou, Yuyin}
% index{Fishman, Elliot}
% index{Yuille, Alan}

%
%\authorrunning{Paper ID 800}
\authorrunning{Y. Xia \textit{et al.}}
% First names are abbreviated in the running head.
% If there are more than two authors, 'et al.' is used.
%
%\institute{Princeton University, Princeton NJ 08544, USA \and
%Springer Heidelberg, Tiergartenstr. 17, 69121 Heidelberg, Germany
%\email{lncs@springer.com}\\
%\url{http://www.springer.com/gp/computer-science/lncs} \and
%ABC Institute, Rupert-Karls-University Heidelberg, Heidelberg, Germany\\
%\email{\{abc,lncs\}@uni-heidelberg.de}}
%
%\institute{Paper ID 800}
\institute{Johns Hopkins University \and
Johns Hopkins Medical Institutions
}
\maketitle              % typeset the header of the contribution
\let\thefootnote\relax\footnote{\text{*} The first two authors equally contributed to the work.}
\begin{abstract}
Pancreatic ductal adenocarcinoma (PDAC) is one of the most lethal cancers among the population. Screening for PDACs in dynamic contrast-enhanced CT is beneficial for early diagnosis. In this paper, we investigate the problem of automated detecting PDACs in multi-phase (arterial and venous) CT scans. Multiple phases provide more information than single phase, but they are unaligned and inhomogeneous in texture,  making it difficult to combine cross-phase information seamlessly. We study multiple phase alignment strategies, \emph{i.e.},  early alignment (image registration), late alignment (high-level feature registration), and slow alignment (multi-level feature registration), and suggest an ensemble of all these alignments as a promising way to boost the performance of PDAC detection. We provide an extensive empirical evaluation on two PDAC datasets and show that the proposed alignment ensemble significantly outperforms previous state-of-the-art approaches, illustrating the strong potential for clinical use.

%present \textit{Deep Alignments}, a deep learning based framework to automatically segment tumor mass in dual-phase (arterial and venous) CT scans. The main challenge of this task is that the phases of CT scans are unaligned and inhomogeneous in texture, making it difficult to combine cross-phase information. To tackle this challenge, we systematically propose and investigate different alignment methods, i.e.. 
%A shared segmentor is appended after the aligned feature maps for accurate labeling. 
%We then find that (i) cross-phase alignments are beneficial for tumor detection (ii)feature-space alignments have better segmentation performance than image registration (iii) different alignment methods are complementary to each other, where an ensemble of the alignment approaches significantly improve the results.
%Then we propose \textit{Deep Alignment Aggregation}, which performs deformable transformations to align and fuse cross-phase information in multi-level feature space.  
%We validate our approach on two PDAC datasets and significantly outperform previous state-of-the-art approaches, 
%in terms of segmentation dice score and tumor sensitivity, 
%illustrating strong potential for clinical use.

\end{abstract}
\section{Introduction}
Pancreatic ductal adenocarcinoma (PDAC) is the third most common cause of cancer death in the US with a dismal five-year survival of merely 9\%~\cite{lucas2019screening}. Computed tomography (CT) is the most widely used imaging modality for the initial evaluation of suspected PDAC. However, due to the subtle early signs of PDACs in CTs, they are easily missed by even experienced radiologists.
%. 

Recently, automated PDAC detection in CT scans based on deep learning has received increasing attention~\cite{chu2019application,chu2019utility,zhu2019multi,zhou2019hyper}, which offers great opportunities in assisting radiologists to diagnosis early-stage PDACs. But, most of these methods only unitize one phase of CT scans, and thus fail to achieve satisfying results.  

\begin{figure}[!t]
\begin{center}
    \includegraphics[width=1.0\linewidth]{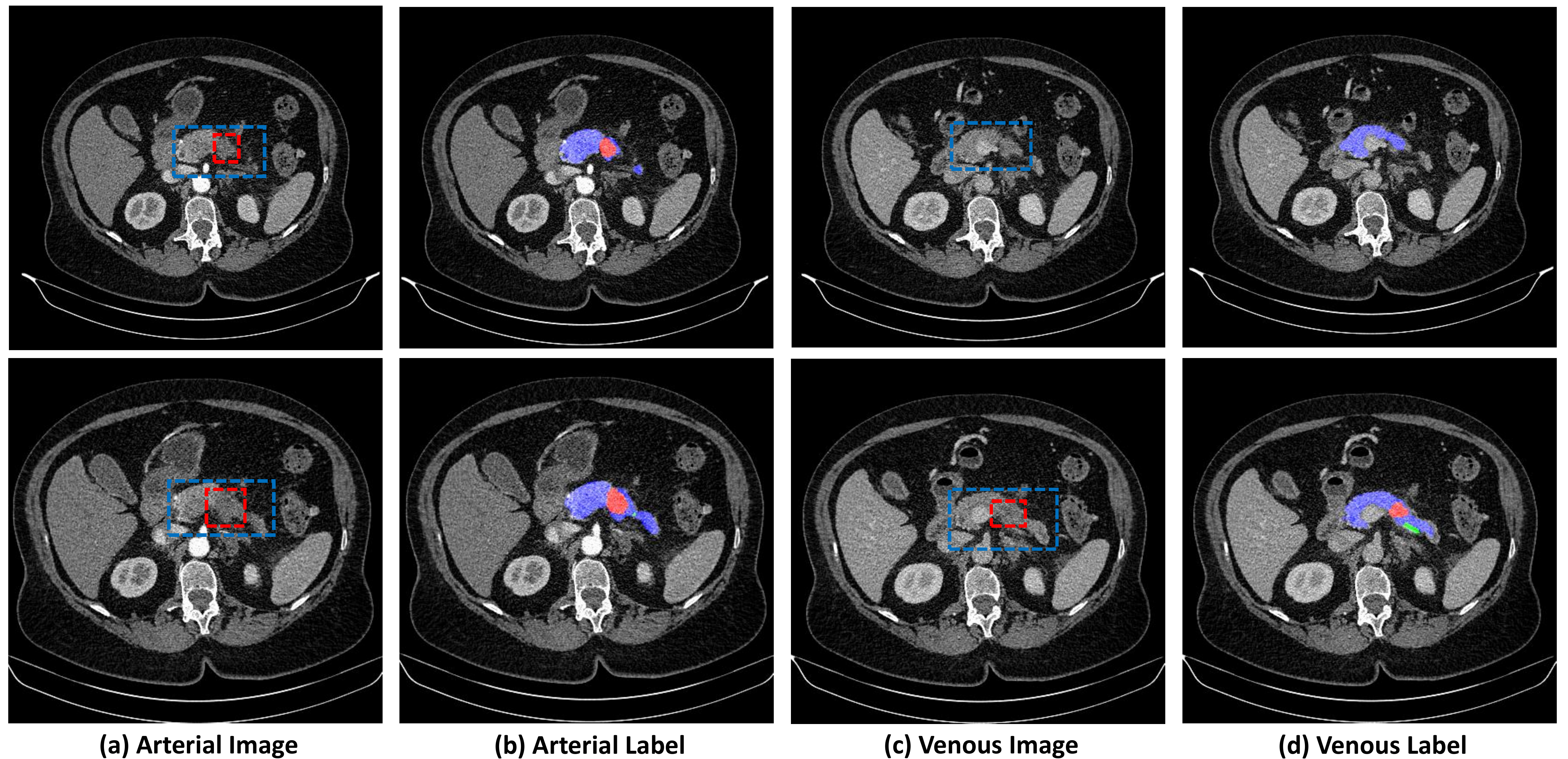}
\end{center}
\caption{
    Visual illustration of opportunity (top row) and challenge (bottom row) for PDAC detection in multi-phase CT scans (normal pancreas tissue - blue, pancreatic duct - green, PDAC mass - red). Top: tumor is barely visible in venous phase alone but more obvious in arterial phase. Bottom: there exist misalignment for images in these two phases given different organ size/shape and image contrast.
}
\label{Fig1}
\end{figure}

In this paper, we aim to develop a deep learning based PDAC detection system taking multiple phases, \emph{i.e.}, arterial and venous, of CT scans into account. This system consists of multiple encoders, each of which encodes information for one phase, and a segmentation decoder, which outputs PDAC detection results. % given by cross-phase information.
Intuitively, multiple phases provide more information than a single phase, which certainly benefits PDAC detection. Nevertheless, how to combine this cross-phase information seamlessly is non-trivial. The challenges lie in two folds: 1) Tumor texture changes are subtle and appear differently across phases; 2) Image contents are not aligned across phases because of inevitable movements of patients during capturing multiple phases of CT scans. Consequently, a sophisticated phase alignment strategy is indispensable for detecting PDAC in multi-phase CT scans. An visual illustration is shown in Fig.~\ref{Fig1}.

We investigate several alignment strategies to combine the information across multiple phases. (1) \textbf{Early alignment} (EA): the alignment can be done in image space by performing image registration between multiple phases; (2) \textbf{Late alignment} (LA): it can be done late in feature space by performing spatial transformation between the encoded high-level features of multiple phases; (3) \textbf{Slow alignment} (SA): it can be also done step-wise in feature space by aggregating multi-level feature transformations between multiple phases. Based on an extensive empirical evaluation on two PDAC datasets~\cite{zhu2019multi,zhou2019hyper}, we observe that 1) All alignment strategies are beneficial for PDAC detection, 2) alignments in feature space leads to better PDAC (tumor) segmentation performance than image registration, and (3) different alignment strategies are complementary to each other, \emph{i.e.}, an ensemble of them (\textbf{Alignment Ensemble}) significantly boosts the results, \emph{e.g.}, approximately 4\% tumor DSC score improvements over our best alignment model.

Our contributions can be summarized as follows:
\begin{itemize}
%\item We provide extensive experimental evaluation of several phase alignment strategies for detecting PDAC in multi-phase CT scans.
\item We propose late and slow alignments as two novel solutions for detecting PDACs in multi-phase CT scans and provide extensive experimental evaluation of different phase alignment strategies.

%We provide extensive experimental evaluation of several phase alignment strategies for detecting PDAC in multi-phase CT scans, among which late and slow alignments are novel solutions to multi-modal segmentation problems.
%\item We highlight an ensemble of early, late and slow alignments as a  promising  way  to  boost  the  performance  of PDAC detection.
\item We highlight early, late and slow alignments are complementary and a simple ensemble of them is a promising way to boost performance of PDAC detection.
\item We validate our approach on two PDAC datasets~\cite{zhu2019multi,zhou2019hyper} and achieve state-of-the-art performances on both of them.
\end{itemize}

\section{Related Work}
\subsubsection{Automated Pancreas and Pancreatic Tumor Segmentation} 
With the recent advances of deep learning, automated pancreas segmentation has achieved tremendous improvements~\cite{roth2015deeporgan,roth2016spatial,cai2016pancreas,zhou2017fixed,yu2018recurrent,zhu20183d,xia2018bridging,dqnpanc}, which is an essential prerequisite for pancreatic tumor detection. Meanwhile, researchers are pacing towards automated detection of pancreatic adenocarcinoma (PDAC), the most common type of pancreatic tumor (85\%)~\cite{ryan2014pancreatic}. Zhu \emph{et al.}~\cite{zhu2019multi} investigated using deep networks to detect PDAC in CT scans but only segmented PDAC masses in venous phase. Zhou \emph{et al.}~\cite{zhou2019hyper} developed the a deep learning based approach for segmenting PDACs in multi-phase CT scans, \emph{i.e.} arterial and venous phase. They used a traditional image registration~\cite{vercauteren2009diffeomorphic} approach for pre-alignment and then applied a deep network that took both phases as input. Different to their method, we also investigate how to register multiple phases in feature space.

\subsubsection{Multi-modal Image Registration and Segmentation} Multi-modal image registration~\cite{roche1998correlation,vercauteren2009diffeomorphic,gaens1998non,deeds} is a fundamental task in medical image analysis. Recently, several deep learning based approaches, motivated by Spatial Transformer Networks~\cite{stn}, are proposed to address this task~\cite{voxelmorph,qin2019multimodelreg,reg2}. In terms of multi-modal segmentation, most of the previous works~\cite{menze2014multimodal,dolz2018hyperdense,zhou2019hyper} perform segmentation on pre-registered multi-modal images. We also study these strategies for multi-modal segmentation, but we explore more, such as variants of end-to-end frameworks that jointly align multiple phases and segment target organs/tissues.

\section{Methodology}
\subsection{Problem Statement}
\label{sec3.1}
We aim at detecting PDACs from unaligned two-phase CT scans, \emph{i.e.}, the venous phase and the arterial phase. Following previous works~\cite{zhou2019hyper,zhu2019multi}, venous phase is our fixed phase and arterial phase is the moving one. For each patient, we have an image  $\mathbf{X}$ and its corresponding label $\mathbf{Y}$ in the venous phase, as well as an arterial phase image $\mathbf{X^\prime}$ without label. The whole dataset is denoted as $S = \{(\mathbf{X_i}, \mathbf{X_i^\prime}, \mathbf{Y_i}) | i = 1,2,..M \}$, where $\mathbf{X_i} \in \mathbb{R}^{H_i\times W_i \times D_i}$, $\mathbf{X_i^\prime} \in \mathbb{R}^{H_i^\prime\times W_i^\prime \times D_i^\prime}$ are 3D volumes representing the two-phase CT scans of the $i$-th patient. $\mathbf{Y_i} \in \mathcal{L}$ is a voxel-wise annotated label map, which have the same $(H_i, W_i, D_i)$ three dimensional size as $\mathbf{X_i}$. Here, $\mathcal{L} = \{0,1,2,3\}$ represents our segmentation targets, \emph{i.e.}, background, healthy pancreas tissue, pancreatic duct (crucial for PDAC clinical diagnoses) and PDAC mass, following previous literature~\cite{zhou2019hyper,zhu2019multi}. Our goal is to find a mapping function $\mathcal{M}$ whose inputs and outputs are a pair of two-phase images $\mathbf{X}, \mathbf{X^\prime}$ and segmentation results $\mathbf{P}$, respectively: $\mathbf{P} = \mathcal{M}(\mathbf{X}, \mathbf{X^\prime})$. The key problem here is how to align $\mathbf{X}$ and $ \mathbf{X^\prime}$, either in image space or feature space.

%\begin{equation}
%\label{eq1}
%    \mathbf{P} = m_\theta(\mathbf{X}, \mathbf{X^\prime})
%\end{equation}

\begin{figure}[!t]
\begin{center}
    \includegraphics[width=1.0\linewidth]{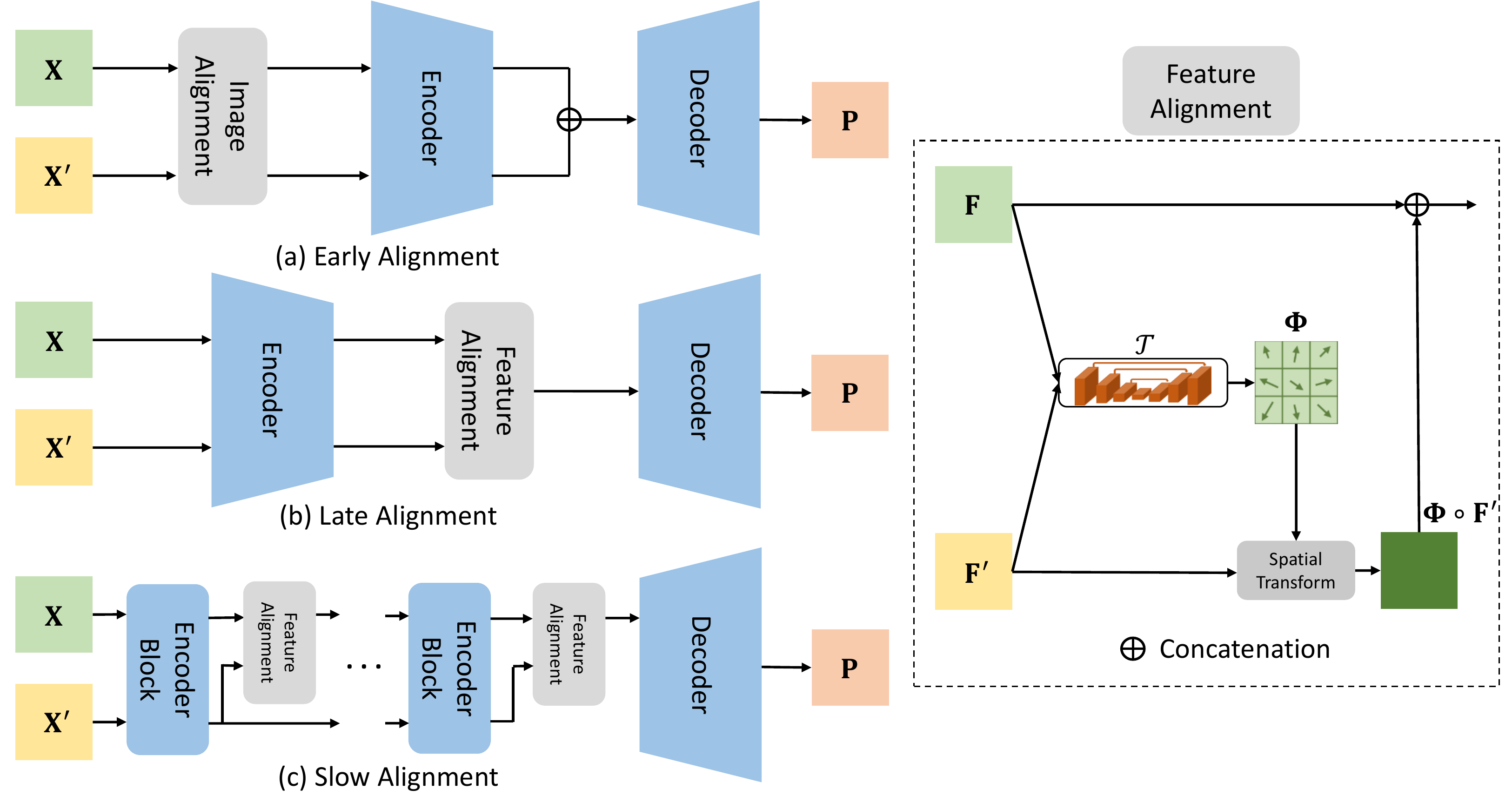}
\end{center}
\caption{
    An illustration of (a) early alignment (image registration) (b) late alignment and (c) slow alignment. Right: feature alignment block.
}
\label{Fig2}
\end{figure}

\subsection{Cross-phase Alignment and Segmentation}
As shown in Fig~\ref{Fig2}, we propose and explore three types of alignment strategies, \emph{i.e.}, early alignment, late alignment and slow alignment, for accurate segmentation.

\subsubsection{Early (image) alignment}
Early alignment, or image alignment strategy is adopted in ~\cite{zhou2019hyper} and some other multi-modal segmentation tasks such as BraTS challenge ~\cite{menze2014multimodal}, where multiple phases (modalities) are first aligned by image registration algorithms and then fed forward into deep networks for segmentation. Here, we utilize a well-known registration algorithm, DEEDS~\cite{deeds}, to estimate the registration field $\mathbf{\Phi}$ from an arterial image $\mathbf{X}^\prime$ to its corresponding venous image $\mathbf{X}$. After registration, we use a network, consisting of two separtae encoders $\mathcal{F}$, $\mathcal{F^\prime}$ and a decoder $\mathcal{G}$, to realize the mapping function $\mathcal{M}$:   

\begin{equation}
\label{eq2}
    \mathbf{P} = \mathcal{M}(\mathbf{X}, \mathbf{X^\prime})=\mathcal{G}(\mathcal{F}(\mathbf{X}) \oplus \mathcal{F^\prime}(\mathbf{\Phi} \circ \mathbf{X^\prime})),
\end{equation}
where $\oplus$ and $\circ$ denote the concatenation of two tensors and the element-wise deformation operations on a tensor, respectively.

This strategy relies on the accuracy of image registration algorithms for information alignment. If such algorithms produce errors, especially possible on subtle texture changes of PDACs, these errors will propagate and there will be no way to rescue (since alignment is only done on image level). Also, it remains a question that how much performance gain a segmentation algorithm will achieve through this separate registration procedure. 

\subsubsection{Late alignment}
An alternative way is late alignment, \emph{i.e.}, alignment in feature space. We first encode the pair of unaligned images $(\mathbf{X},\mathbf{X}^\prime)$ with two phase-specific encoders $(\mathcal{F}, \mathcal{F}^\prime)$, respectively. The encoded features of the two images, \emph{i.e.}, $\mathbf{F} = \mathcal{F}(\mathbf{X})$ and $\mathbf{F^\prime} = \mathcal{F}^\prime(\mathbf{X^\prime})$, are presumablely in a shared feature space.
%Motivated by Spatial Transformer Network (STN)~\cite{stn} and its application to image registration~\cite{voxelmorph}, 
We then use a network $\mathcal{T}$ to estimate the deformable transformation field $\mathbf{\Phi}$ from arterial (moving) to venous (fixed) in the feature space by $\mathbf{\Phi} = \mathcal{T}(\mathbf{F}, \mathbf{F^\prime})$. We apply the estimated transformation field $\mathbf{\Phi}$ to feature map $\mathbf{F}^\prime$, then concatenate this transformed feature map $\mathbf{\Phi} \circ \mathbf{F}^\prime$ to $\mathbf{F}$. The segmentation result $\mathbf{P}$ is obtained by feeding the concatenation to a decoder $\mathcal{G}$:

\begin{equation}
    \mathbf{P} = \mathcal{M}(\mathbf{X}, \mathbf{X^\prime})=\mathcal{G}(\mathbf{F} \oplus \mathbf{\Phi} \circ \mathbf{F^\prime})= \mathcal{G}(\mathcal{F}(\mathbf{X}) \oplus \mathcal{T}(\mathbf{F}, \mathbf{F^\prime}) \circ \mathcal{F}^\prime(\mathbf{X^\prime})).
\end{equation}

We name such operation as ``late alignment" since the alignment is performed at the last block of feature encoders. 

\subsubsection{Slow alignment} Late alignment performs one-off registration between two phases by only using high level features. However, it is known that the low level features of the deep network contain more image details, which motivates us to gradually align and propagate the features from multiple levels of the deep network. Following this spirit, we propose slow alignment, which leverages a stack of convolutional encoders and feature alignment blocks to iteratively align feature maps of two phases. 

Let $k$ be an integer which is not less than 1 and ($\mathbf{F}_{k-1}$, $\mathbf{F}_{k-1}^\prime$) are the fused (aligned to the venous phase) feature map and the arterial feature map outputted by the $(k-1)^{th}$ convolutional encoder, respectively. First, they are encoded by a pair of convolutional encoders ($\mathcal{F}_k$, $\mathcal{F}_k^\prime$), respectively, which results in the venous feature map  $\mathbf{F}_k=\mathcal{F}_k(\mathbf{F}_{k-1})$ and the arterial feature map $\mathbf{F}_k^\prime=\mathcal{F}_k^\prime(\mathbf{F}_{k-1}^\prime)$ at the $k$-th layer. Then a feature alignment block estimates a transformation field from the arterial (moving) phase to the venous (fixed) phase by 
\begin{equation}
    \mathbf{\Phi}_k = \mathcal{T}_k(\mathcal{F}_k(\mathbf{F}_{k-1}), \mathcal{F}_k^\prime(\mathbf{F}_{k-1}^\prime)),
\end{equation}
 where $\mathcal{T}_k$ is a small U-Net. We apply the transformation field $\mathbf{\Phi}_k$ to the arterial (moving) phase, resulting in transformed arterial feature map $\mathbf{\Phi}_k \circ \mathcal{F}_k^\prime(\mathbf{F}_{k-1}^\prime)$. Finally, the transformed arterial feature map is concatenated with the venous feature map $\mathcal{F}_k(\mathbf{F}_{k-1})$, resulting in the fused feature map at the $k^{th}$ layer: 

\begin{equation}
        \mathbf{F}_k = \mathcal{F}_k(\mathbf{F}_{k-1}) \oplus \mathbf{\Phi}_k \circ \mathcal{F}_k^\prime(\mathbf{F}_{k-1}^\prime).
\end{equation}
Let us rewrite the above process by a function $\mathcal{R}_k$: $\mathbf{F}_{k} = \mathcal{R}_{k}(\mathbf{F}_{k-1}, \mathbf{F}_{k-1}^\prime)$ and define $\mathbf{F}_0=\mathbf{X}$ and $\mathbf{F}_0^\prime=\mathbf{X}^\prime$, then we can iteratively derive the fused feature map at $n$-th convolutional encoder:
\begin{align}
\mathbf{F}_n = \mathcal{R}_n \bigg(\mathcal{R}_{n-1} \Big(\cdots \big(\mathcal{R}_1(\mathbf{F}_0, \mathbf{F}_0^{\prime}),\mathbf{F}_{1}^\prime\big),\cdots\Big), \mathbf{F}_{n-1}^\prime\bigg),    
\end{align}
where $\mathbf{F}_{n-1}^\prime = \mathcal{F}_{n-1}^\prime(\mathcal{F}_{n-2}^\prime(\cdots(\mathcal{F}_1^\prime(\mathbf{F}_0^{\prime})))$.
The final fused feature map $\mathbf{F}_n$ is fed to the decoder $\mathcal{G}$ to compute the segmentation result $\mathbf{P}$:
\begin{equation}
\label{eq:sa}
\mathbf{P} = \mathcal{M}(\mathbf{X}, \mathbf{X}^\prime)=\mathcal{G}(\mathbf{F}_n).
\end{equation}
%We combine such one step of encoding and alignment as a whole function $\mathcal{R}_k$ for convenience:
%\begin{equation}
%    \begin{aligned}
%        \mathbf{F_{k}} = \mathcal{R}_{k}(\mathbf{F_{k-1}}, %\mathbf{F_{k-1}^\prime})
%    \end{aligned}
%\end{equation}

%Slow alignment is a $n$-stack of convolutional encoders and feature alignment block. Same as late alignment, we also apply a segmentation decoder $\mathcal{G}$ to compute final segmentation masks. Mathematically, 

% \begin{equation}
% \label{eq:sa}
%          \mathbf{P} = \mathcal{M}(\mathbf{X}, \mathbf{X}^\prime) = \mathcal{G}(\mathbf{F}_n) = \mathcal{G}(\mathcal{R}_n (\mathcal{R}_{n-1} (\cdot \cdot \cdot (\mathcal{R}_1(\mathbf{X}, \mathbf{X^\prime}))))
% \end{equation}

%where $\mathbf{F}_n$ and $\mathbf{F}_n^\prime$ are the final encoded feature of venous and arterial phase, respectively. 
%In practice, each encoder function $f_k$ is implemented by a residual block~\cite{he2016deep} and each transformation field estimation function $\phi_k$ is implemented by a shallow and light-weighted U-Net~\cite{ronneberger2015u}. 
 %$\mathbf{P} = \mathcal{G}(\mathbf{F_n})$. ($\mathbf{F_n}$ has been already fused with arterial deformed feature.)

\subsubsection{Alignment Ensemble} We ensemble the three proposed alignment variants by simple majority voting of the predictions. The goal of the ensemble are in two folds, where the first is to improve overall performance and the second is to see whether these three alignment methods are complementary. Usually, an ensemble of complementary approaches can lead to large improvements. 

% \subsection{Segmentation}
% After feature alignment, we append a shared segmentation network $g(\cdot)$ which aims at predicting segmentation labels from the encoded features, where we applied a same decoder for both phases. This design not only decreases the number of parameters of the overall model, but can also encourage the encoded feature $F_n^A$ and $F_n^B$ to be more similar, which potentially benefits the feature combination of the two phases. Finally, the output segmentation masks of both phases can be computed by:

% \begin{equation}
%     \begin{aligned}
%         \mathbf{P^A} = g(\mathbf{F_n^A}) \\
%         \mathbf{P^B} = g(\mathbf{F_n^B})
%     \end{aligned}
% \end{equation}

\section{Experiments and discussion}
\subsection{Dataset and evaluation}
We evaluate our approach on two PDAC datasets, proposed in ~\cite{zhu2019multi} and ~\cite{zhou2019hyper} respectively. For the ease of presentation, we regard the former as PDAC dataset \Romannum{1} and the latter as PDAC dataset \Romannum{2}. PDAC dataset \Romannum{1} contains 439 CT scans in total, in which 136 cases are diagnosed with PDAC and 303 cases are normal. Annotation contains voxel-wise labeled pancreas and PDAC mass. Evaluation is done by 4 fold cross-validation on these cases following~\cite{zhu2019multi}. PDAC dataset \Romannum{2} contains 239 CT scans, all from PDAC patients, with pancreas, pancreatic duct (crucial for PDAC detection) and PDAC mass annotated. Evaluation are done by 3 fold cross-validation following~\cite{zhou2019hyper}. 

All cases contain two phases: arterial phase and venous phase, with a spacing of 0.5mm in axial view and all annotations are verified by experienced board certified radiologists. The segmentation accuracy is evaluated using the Dice-S\o rensen coefficient (DSC): $\mathrm{DSC}$ $(\mathcal{Y},\mathcal{Z}) = \frac{2\times|\mathcal{Y}\cap \mathcal{Z}|}{|\mathcal{Y}|+|\mathcal{Z}|}$, which has a range of $[0, 1]$ with 1 implying a perfect prediction for each class. On dataset \Romannum{1}, we also evaluate classification accuracy by sensitivity and specificity following a ``segmentation for classification" strategy proposed in~\cite{zhu2019multi}.

\subsection{Implementation details}
We implemented our network with PyTorch. The CT scans are first truncated within a range of HU value [-100, 240] and normalized with zero mean and unit variance. In training stage, we randomly crop a patch size of $96^3$ in roughly the same position from both arterial and venous phases. The optimization objective is Dice loss~\cite{vnet}. We use SGD optimizer with initial learning 0.005 and a cosine learning rate schedule for 40k iterations. For all our experiments, we implement the encoder and decoder architecture as U-Net~\cite{ronneberger2015u} with 4 downsampling layers, making a total alignments of $n=4$ in Eq~\ref{eq:sa}. The transformation fields are estimated by light-weighted U-Nets in late alignment and slow alignment, each is $\sim$8$\times$ smaller than the large U-Net for segmentation, since the inputs of the small U-Nets are already the compact encoded features. The computation of EA/LA/SA is approximately 1.5/1.7/1.9 times of the computation of a single-phase U-Net. The image registration algorithm for our early alignment is DEEDS~\cite{deeds}.

\begin{table}[!t]
\centering
\begin{tabular}{l|c|c|c|c|c|c}    
\hline
Method & N.Pancreas  & A.Pancreas & Tumor & Misses & Sens. & Spec.\\
\hline
U-Net~\cite{ronneberger2015u}  & 86.9$\pm$8.6  & 81.0$\pm$10.8 & 57.3$\pm$28.1& 10/136& 92.7&\textbf{99.0}\\
V-Net~\cite{vnet}  & 87.0$\pm$8.4  & 81.6$\pm$10.2& 57.6$\pm$27.8& 11/136& 91.9&99.0\\
MS C2F~\cite{zhu2019multi} &  84.5$\pm$ 11.1 & 78.6 $\pm$ 13.3 & 56.5$\pm$ 27.2 & 8/136 &  94.1& 98.5\\
%HPN-ResDSN~\cite{zhou2019hyper} & & & & & & \\
\hline
Baseline - NA & 85.8$\pm$8.0 & 79.5$\pm$11.2 & 58.4$\pm$27.4 & 11/136& 91.9& 96.0 \\
\hline
Ours - EA &  86.7$\pm$9.7 & 81.8$\pm$10.0 & 60.9$\pm$26.5& 4/136& 97.1& 94.5\\
Ours - LA &  87.5$\pm$7.6 & 82.0$\pm$10.3 & 62.0$\pm$27.0 & 7/136& 94.9& 96.0\\
Ours - SA &  87.0$\pm$7.8 & 82.8$\pm$9.4 & 60.4$\pm$27.4 & 4/136& 97.1& 96.5\\
\hline
Ours - Ensemble &  \textbf{87.6$\pm$7.8} & \textbf{83.3$\pm$8.2} & \textbf{64.4$\pm$25.6} & \textbf{4/136}& \textbf{97.1}& 96.0\\
\hline
\end{tabular}
\caption{
    Results on PDAC dataset \Romannum{1} with both healthy and pathological cases. We compare our variants of alignment methods with the state-of-the-art method~\cite{zhu2019multi} as well as our baseline - no align (NA) version. ``Misses" represents the number of cases failed in tumor detection. We also report healthy vs. pathological case classification (sensitivity and specificity) based on segmentation results. The last row is the ensemble of the three alignments.
}
\label{Tab1}
\end{table}
% 60.5$\pm$27.4

\subsection{Results}
%\subsubsection{PDAC dataset \Romannum{1}} Results on dataset \Romannum{1} is summerized in table~\ref{Tab1}

%\subsubsection{PDAC dataset \Romannum{2}} Results on dataset \Romannum{2} is summerized in table~\ref{Tab2}
Results on dataset \Romannum{1} and \Romannum{2} are summarized in Table~\ref{Tab1} and Table~\ref{Tab2} respectively, where our approach achieves the state-of-the-art performance on both datasets. Based on the results, we have three observations which leads to three findings.

\textbf{Dual-phase alignments are beneficial for detecting PDACs in multi-phase CT scans.} On both datasets, our approaches, \emph{i.e.} early alignment, late alignment and slow alignment, outperform single phase algorithms, i.e. U-Net~\cite{ronneberger2015u}, V-Net~\cite{vnet}, ResDSN~\cite{zhu20183d} and MS C2F~\cite{zhu2019multi}, as well as our non-alignment dual-phase version (Baseline-NA).

\begin{table}[!t]
\centering
\begin{tabular}{l|c|c|c|c}    
\hline
Method  & A.Pancreas  & Tumor & Panc. duct& Misses \\
\hline
U-Net~\cite{ronneberger2015u}   & 79.61$\pm$10.47 & 53.08$\pm$27.06 & 40.25$\pm$27.89 & 11/239\\
ResDSN~\cite{zhu20183d}   & 84.92$\pm$7.70 & 56.86$\pm$26.67 & 49.81$\pm$26.23 & 11/239 \\
HPN-U-Net~\cite{zhou2019hyper}   & 82.45$\pm$9.98 & 54.36$\pm$26.34 & 43.27$\pm$26.33& -/239\\
HPN-ResDSN~\cite{zhou2019hyper}   & 85.79$\pm$8.86 & 60.87$\pm$24.95 & 54.18$\pm$24.74 & 7/239 \\
\hline
Ours - EA & 83.65$\pm$9.22 & 60.87$\pm$22.15 & 55.38$\pm$29.47 & \textbf{5/239} \\
Ours - LA  & 86.82$\pm$6.13 & 62.02$\pm$24.53 & 64.35$\pm$29.94 & 9/239 \\
Ours - SA & 87.13$\pm$5.85 & 61.24$\pm$24.26 & 64.19$\pm$29.46 & 8/239  \\
\hline
Ours - Ensemble & \textbf{87.37$\pm$5.67} & \textbf{64.14$\pm$21.16} & \textbf{64.38$\pm$29.67} & 6/239 \\
\hline
\end{tabular}
\caption{
    Results on PDAC dataset \Romannum{2} with pathological cases only. We compare our variants of alignment methods with the state-of-the-art method~\cite{zhou2019hyper}. ``Misses" represents the number of cases failed in tumor detection. The last row is the ensemble of the three alignments.
}
\label{Tab2}
\end{table}

\begin{figure}[!b]
\begin{center}
    \includegraphics[width=1.0\linewidth]{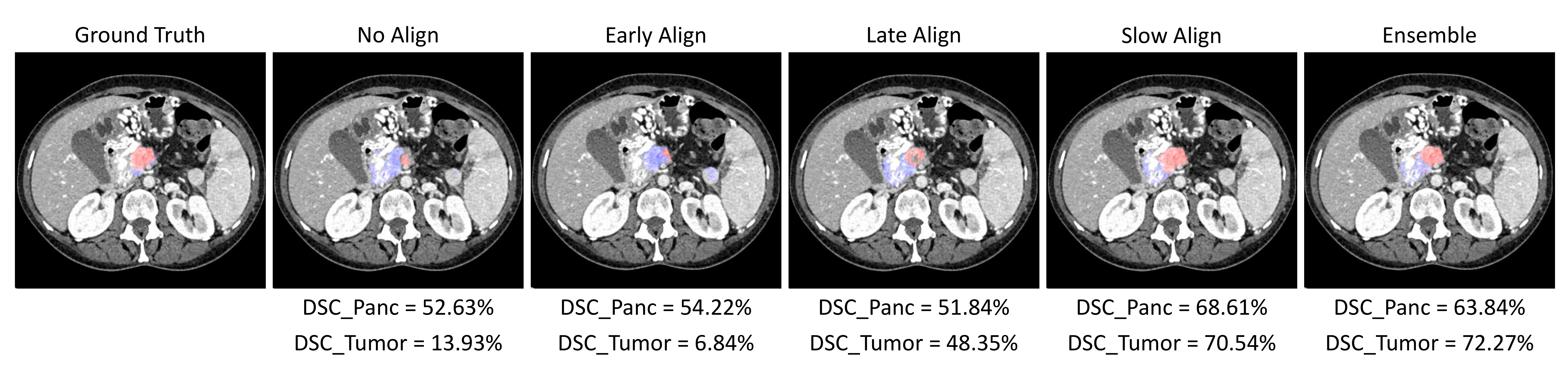}
\end{center}
\caption{
    An example of PDAC dataset \Romannum{1} on venous phase. From left to right, we display ground-truth, prediction of our baseline without alignment, prediction of our early align, late align, slow align and alignment ensemble. Our feature space alignments (LA, SA) outperform no-align baseline and image registration (EA). Ensemble of the three alignment predictions also improves tumor segmentation DSC score. 
}
\label{Fig3}
\end{figure}

\textbf{Feature space alignments have larger improvements on segmentation performances than early alignments.} Generally speaking for both datasets, our feature space alignment models (LA, SA) outperform image registration based approaches, i.e. HPN, Ours-EA, in terms of  segmentation performance. Since early alignment methods apply image registration in advance, they do not guarantee a final improvement on segmentation performance. In contrast, feature space alignment methods jointly align and segment the targets in an end-to-end fashion by optimizing the final segmentation objective function, which leads to a larger improvements compared with single phase or naive dual phase methods without alignment. However, we indeed observe that early alignment leads to relatively less false negatives (misses).

\textbf{An ensemble of the three alignment strategies significantly improve the performances.} For both dataset, Ours-Ensemble achieves the best performances, illustrating that the three alignment strategies are \textbf{complementary} to each other. An ensemble leads to significant performance gain (relatively 4\% improvements on tumor segmentation DSC score compared to the best alignment model from 62.0\% to 64.4\%) and achieves the state-of-the-art performances on both datasets. A qualitative analysis is also shown in Fig~\ref{Fig3}.

Last but not least, our alignment approaches also improve the sensitivity of healthy vs. pathological classification. In dataset \Romannum{1}, we adopt the same ``segmentation for classification" strategy as in~\cite{zhou2019hyper}, which classifies a case as pathological if we are able to detect any tumor mass larger than 50 voxels. Our approach can improve the overall sensitivity from 94.1\% to 97.1\% by reducing misses from 8 to 4, which is beneficial for the early detection of PDAC. Our approach thus has valuable potential of winning precious time for early treatments for patients. 

\section{Conclusion}
In this paper, we study three types of alignment approaches for detecting pancreatic adenocarcinoma (PDACs) in multi-phase CT scans. Early alignment first applies registration in image space and then segment with a deep network. Late alignment and slow alignment jointly align and segment with an end-to-end deep network. The former aligns in the final encoded feature space while the latter aligns multi-stage features and propagate slowly. An ensemble of the three approaches improve the performances significantly illustrating these alignment variants are complementary to each other. We achieve the state-of-the-art performances on two PDAC datasets.

\noindent\textbf{Acknowledgement} This work was supported by the Lustgarten Foundation for Pancreatic Cancer Research and also
supported by NSFC No. 61672336.

%Another type of pancreatic tumor, pancreatic neuroendocrine tumor (PanNet), 

%Our main contributions are summarized as follows:
%\begin{itemize}
%  \item We investigate the problem of pancreatic tumor segmentation in contrast-enhanced dynamic CT scans. We propose a novel and effective network, namely \textit{MorphSegNet}, which takes both arterial and venous phase CT scans as input to segment pancreas, pancreatic duct and pancreatic tumor, in an end-to-end fashion. 
%  \item We propose dual-way feature-space transformation module to incorporate unaligned cross-phase cues for accurate segmentation. This design makes our framework free of pre-registration between phases, and thus significantly improves testing speed. In our framework, such module is densely stacked to automatically align multi-stage features, leading to improved segmentation results.
%  \item Our team collects a dataset of multi-phase CT scans that contains xxx PDAC patients, xxx PNET patients and xxx normal patients with voxel-wise annotations. To the best of our knowledge, this is the largest and best-quality dataset aiming at pancreatic cancer detection. Our approach outperform other approaches in the measurement of sensitivity, specificity and segmentation dice score. 
%\end{itemize}
%
% ---- Bibliography ----
%
% BibTeX users should specify bibliography style 'splncs04'.
% References will then be sorted and formatted in the correct style.
%
\bibliographystyle{splncs04}
\bibliography{ref}

\begin{thebibliography}{10}
\providecommand{\url}[1]{\texttt{#1}}
\providecommand{\urlprefix}{URL }
\providecommand{\doi}[1]{https://doi.org/#1}

\bibitem{voxelmorph}
Balakrishnan, G., Zhao, A., Sabuncu, M.R., Guttag, J., Dalca, A.V.: Voxelmorph:
  a learning framework for deformable medical image registration. IEEE
  transactions on medical imaging  (2019)

\bibitem{cai2016pancreas}
Cai, J., Lu, L., Zhang, Z., Xing, F., Yang, L., Yin, Q.: Pancreas segmentation
  in mri using graph-based decision fusion on convolutional neural networks.
  In: International Conference on Medical Image Computing and Computer-Assisted
  Intervention. pp. 442--450. Springer (2016)

\bibitem{chu2019utility}
Chu, L.C., Park, S., Kawamoto, S., Fouladi, D.F., Shayesteh, S., Zinreich,
  E.S., Graves, J.S., Horton, K.M., Hruban, R.H., Yuille, A.L., et~al.: Utility
  of ct radiomics features in differentiation of pancreatic ductal
  adenocarcinoma from normal pancreatic tissue. American Journal of
  Roentgenology  \textbf{213}(2),  349--357 (2019)

\bibitem{chu2019application}
Chu, L.C., Park, S., Kawamoto, S., Wang, Y., Zhou, Y., Shen, W., Zhu, Z., Xia,
  Y., Xie, L., Liu, F., et~al.: Application of deep learning to pancreatic
  cancer detection: Lessons learned from our initial experience. Journal of the
  American College of Radiology  \textbf{16}(9),  1338--1342 (2019)

\bibitem{dolz2018hyperdense}
Dolz, J., Gopinath, K., Yuan, J., Lombaert, H., Desrosiers, C., Ayed, I.B.:
  Hyperdense-net: a hyper-densely connected cnn for multi-modal image
  segmentation. IEEE transactions on medical imaging  \textbf{38}(5),
  1116--1126 (2018)

\bibitem{gaens1998non}
Gaens, T., Maes, F., Vandermeulen, D., Suetens, P.: Non-rigid multimodal image
  registration using mutual information. In: International Conference on
  Medical Image Computing and Computer-Assisted Intervention. pp. 1099--1106.
  Springer (1998)

\bibitem{deeds}
Heinrich, M.P., Jenkinson, M., Brady, M., Schnabel, J.A.: Mrf-based deformable
  registration and ventilation estimation of lung ct. IEEE transactions on
  medical imaging  \textbf{32}(7),  1239--1248 (2013)

\bibitem{stn}
Jaderberg, M., Simonyan, K., Zisserman, A., et~al.: Spatial transformer
  networks. In: Advances in neural information processing systems. pp.
  2017--2025 (2015)

\bibitem{lucas2019screening}
Lucas, A.L., Kastrinos, F.: Screening for pancreatic cancer. Jama
  \textbf{322}(5),  407--408 (2019)

\bibitem{dqnpanc}
Man, Y., Huang, Y., Feng, J., Li, X., Wu, F.: Deep q learning driven ct
  pancreas segmentation with geometry-aware u-net. IEEE transactions on medical
  imaging  \textbf{38}(8),  1971--1980 (2019)

\bibitem{menze2014multimodal}
Menze, B.H., Jakab, A., Bauer, S., Kalpathy-Cramer, J., Farahani, K., Kirby,
  J., Burren, Y., Porz, N., Slotboom, J., Wiest, R., et~al.: The multimodal
  brain tumor image segmentation benchmark (brats). IEEE transactions on
  medical imaging  \textbf{34}(10),  1993--2024 (2014)

\bibitem{vnet}
Milletari, F., Navab, N., Ahmadi, S.A.: V-net: Fully convolutional neural
  networks for volumetric medical image segmentation. In: 2016 Fourth
  International Conference on 3D Vision (3DV). pp. 565--571. IEEE (2016)

\bibitem{qin2019multimodelreg}
Qin, C., Shi, B., Liao, R., Mansi, T., Rueckert, D., Kamen, A.: Unsupervised
  deformable registration for multi-modal images via disentangled
  representations. In: International Conference on Information Processing in
  Medical Imaging. pp. 249--261. Springer (2019)

\bibitem{roche1998correlation}
Roche, A., Malandain, G., Pennec, X., Ayache, N.: The correlation ratio as a
  new similarity measure for multimodal image registration. In: International
  Conference on Medical Image Computing and Computer-Assisted Intervention. pp.
  1115--1124. Springer (1998)

\bibitem{ronneberger2015u}
Ronneberger, O., Fischer, P., Brox, T.: U-net: Convolutional networks for
  biomedical image segmentation. In: International Conference on Medical image
  computing and computer-assisted intervention. pp. 234--241. Springer (2015)

\bibitem{roth2015deeporgan}
Roth, H.R., Lu, L., Farag, A., Shin, H.C., Liu, J., Turkbey, E.B., Summers,
  R.M.: Deeporgan: Multi-level deep convolutional networks for automated
  pancreas segmentation. In: International conference on medical image
  computing and computer-assisted intervention. pp. 556--564. Springer (2015)

\bibitem{roth2016spatial}
Roth, H.R., Lu, L., Farag, A., Sohn, A., Summers, R.M.: Spatial aggregation of
  holistically-nested networks for automated pancreas segmentation. In:
  International Conference on Medical Image Computing and Computer-Assisted
  Intervention. pp. 451--459. Springer (2016)

\bibitem{ryan2014pancreatic}
Ryan, D.P., Hong, T.S., Bardeesy, N.: Pancreatic adenocarcinoma. New England
  Journal of Medicine  \textbf{371}(11),  1039--1049 (2014)

\bibitem{vercauteren2009diffeomorphic}
Vercauteren, T., Pennec, X., Perchant, A., Ayache, N.: Diffeomorphic demons:
  Efficient non-parametric image registration. NeuroImage  \textbf{45}(1),
  S61--S72 (2009)

\bibitem{xia2018bridging}
Xia, Y., Xie, L., Liu, F., Zhu, Z., Fishman, E.K., Yuille, A.L.: Bridging the
  gap between 2d and 3d organ segmentation with volumetric fusion net. In:
  International Conference on Medical Image Computing and Computer-Assisted
  Intervention. pp. 445--453. Springer (2018)

\bibitem{yu2018recurrent}
Yu, Q., Xie, L., Wang, Y., Zhou, Y., Fishman, E.K., Yuille, A.L.: Recurrent
  saliency transformation network: Incorporating multi-stage visual cues for
  small organ segmentation. In: Proceedings of the IEEE Conference on Computer
  Vision and Pattern Recognition. pp. 8280--8289 (2018)

\bibitem{zhou2019hyper}
Zhou, Y., Li, Y., Zhang, Z., Wang, Y., Wang, A., Fishman, E.K., Yuille, A.L.,
  Park, S.: Hyper-pairing network for multi-phase pancreatic ductal
  adenocarcinoma segmentation. In: International Conference on Medical Image
  Computing and Computer-Assisted Intervention. pp. 155--163. Springer (2019)

\bibitem{zhou2017fixed}
Zhou, Y., Xie, L., Shen, W., Wang, Y., Fishman, E.K., Yuille, A.L.: A
  fixed-point model for pancreas segmentation in abdominal ct scans. In:
  International Conference on Medical Image Computing and Computer-Assisted
  Intervention. pp. 693--701. Springer (2017)

\bibitem{reg2}
Zhu, W., Myronenko, A., Xu, Z., Li, W., Roth, H., Huang, Y., Milletari, F., Xu,
  D.: Neurreg: Neural registration and its application to image segmentation.
  In: The IEEE Winter Conference on Applications of Computer Vision. pp.
  3617--3626 (2020)

\bibitem{zhu20183d}
Zhu, Z., Xia, Y., Shen, W., Fishman, E., Yuille, A.: A 3d coarse-to-fine
  framework for volumetric medical image segmentation. In: 2018 International
  Conference on 3D Vision (3DV). pp. 682--690. IEEE (2018)

\bibitem{zhu2019multi}
Zhu, Z., Xia, Y., Xie, L., Fishman, E.K., Yuille, A.L.: Multi-scale
  coarse-to-fine segmentation for screening pancreatic ductal adenocarcinoma.
  In: International Conference on Medical Image Computing and Computer-Assisted
  Intervention. pp. 3--12. Springer (2019)

\end{thebibliography}
%
%\bibliography{ref}

\end{document}